\documentclass{ws-ijmpb}

\begin{document}

\markboth{A.M. Ettouhami}
{Positional order of the pinned Abrikosov vortex lattice in samples of finite thickness}

%
\catchline{}{}{}{}{}
%

\title{
Positional order of the pinned Abrikosov flux-line lattice \\
in samples of finite thickness
}
\author{A.M. Ettouhami}

\address{Department of Physics, University of Florida, P.O. Box 118440, 
Gainesville, FL 32606, USA
\\
mouneim@phys.ufl.edu}

\maketitle

\begin{history}
\end{history}

\begin{abstract}

We study translational correlations of the vortex {\em center of mass} positions of the Abrikosov 
flux-line lattice in superconducting samples of finite thickness $L$ (along the direction of 
flux-lines). The Larkin correlation lengths for the center of mass mode of the flux lines in the 
presence of point and correlated disorder are computed, and we find that in the case of point 
disorder the average (i.e. center of mass) position of flux lines maintains positional order on 
length scales which scale like $\sim\sqrt{L}$ in $2+1$ dimensions. 
On still longer length scales, however, we find using a replica 
Gaussian variational approach that center of mass correlations cross over to a power law growth 
of the form $r_\perp/L$, which should be observable in superconducting thin films.
 
\end{abstract}


\section{Introduction}
\label{lro-intro}

There has been a great interest in the properties of the Abrikosov flux-line
lattice\cite{Abrikosov} (FLL) in high temperature superconductors\cite{Bednorz-Muller} 
(HTSCs) during the past 
eighteen years. This interest was motivated by the remarkable physical properties of these materials,
such as their high superconducting critical temperatures and their anisotropic, layered structure, 
with both properties dramatically increasing the importance of thermal fluctuations. Thus, it has 
quickly been recognized that HTSCs are an excellent system for studying the combined effects of 
thermal fluctuations and quenched disorder on flux-line assemblies. These combined effects lead 
to a remarkably rich phase diagram for the flux-line system, the phenomenology of which, even 
after many years of experimental and theoretical investigations, continues to pose many 
exciting and challenging questions.\cite{Blatter-et-al,Nattermann-Scheidl}

\begin{figure}[t]
\includegraphics[scale=0.6]{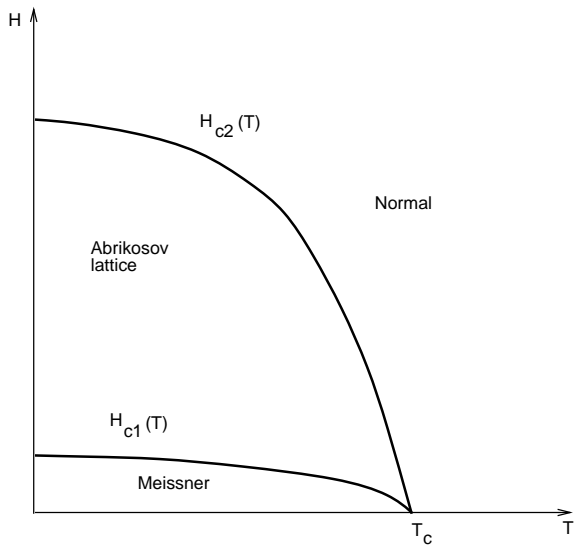}
\caption[]{
Phase diagram of the flux line system in conventional (low-temperature) superconductors. Below $H_{c1}(T)$,
the superconductor is in the so-called Meissner state, characterized by perfect diamagnetism,
while above $H_{c2}(T)$ the system becomes normal (no superconductivity).
}\label{Fig_Abrikosov1}
\end{figure}

\begin{figure}[h]
\includegraphics[scale=0.6]{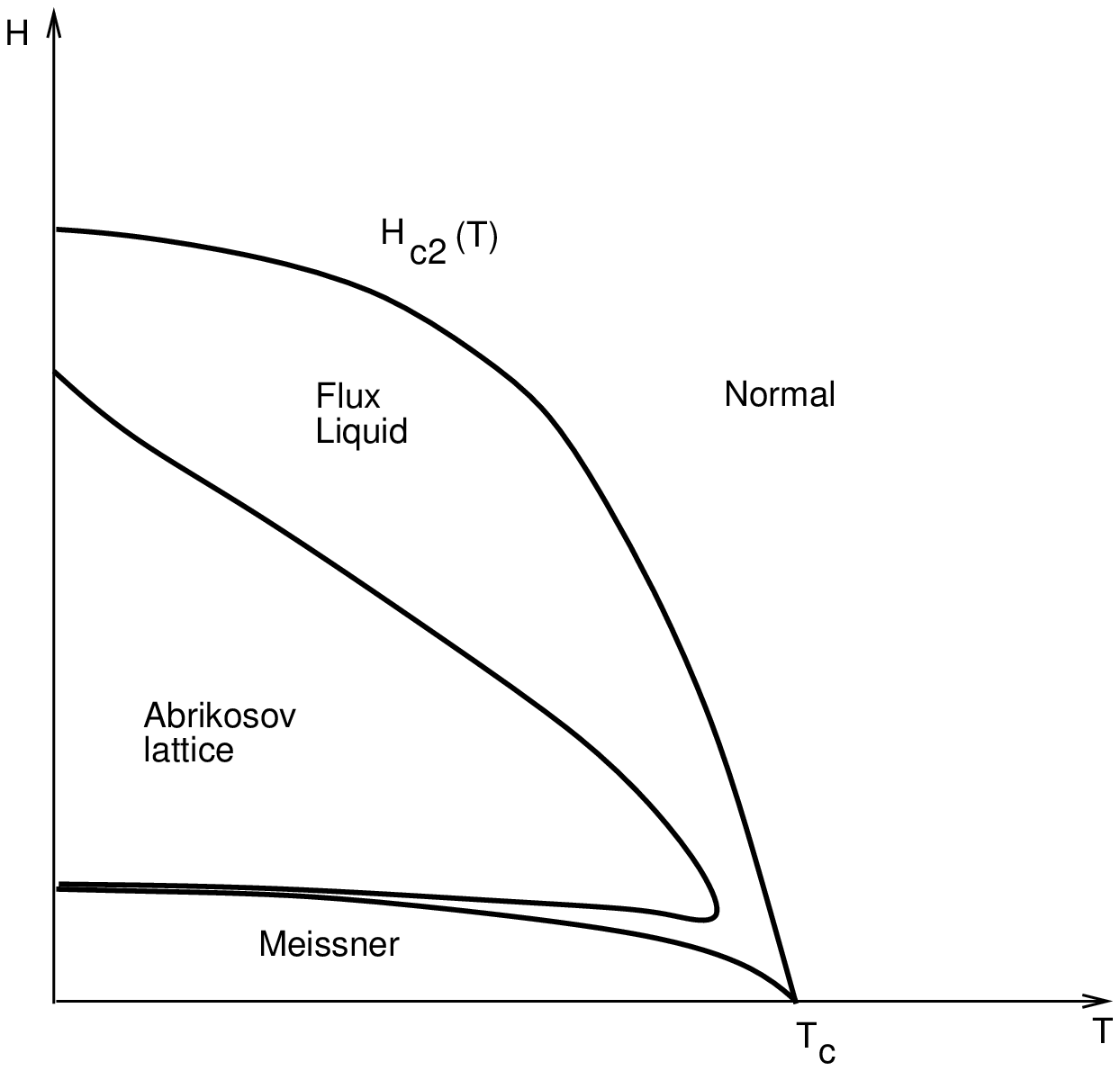}
\includegraphics[scale=0.6]{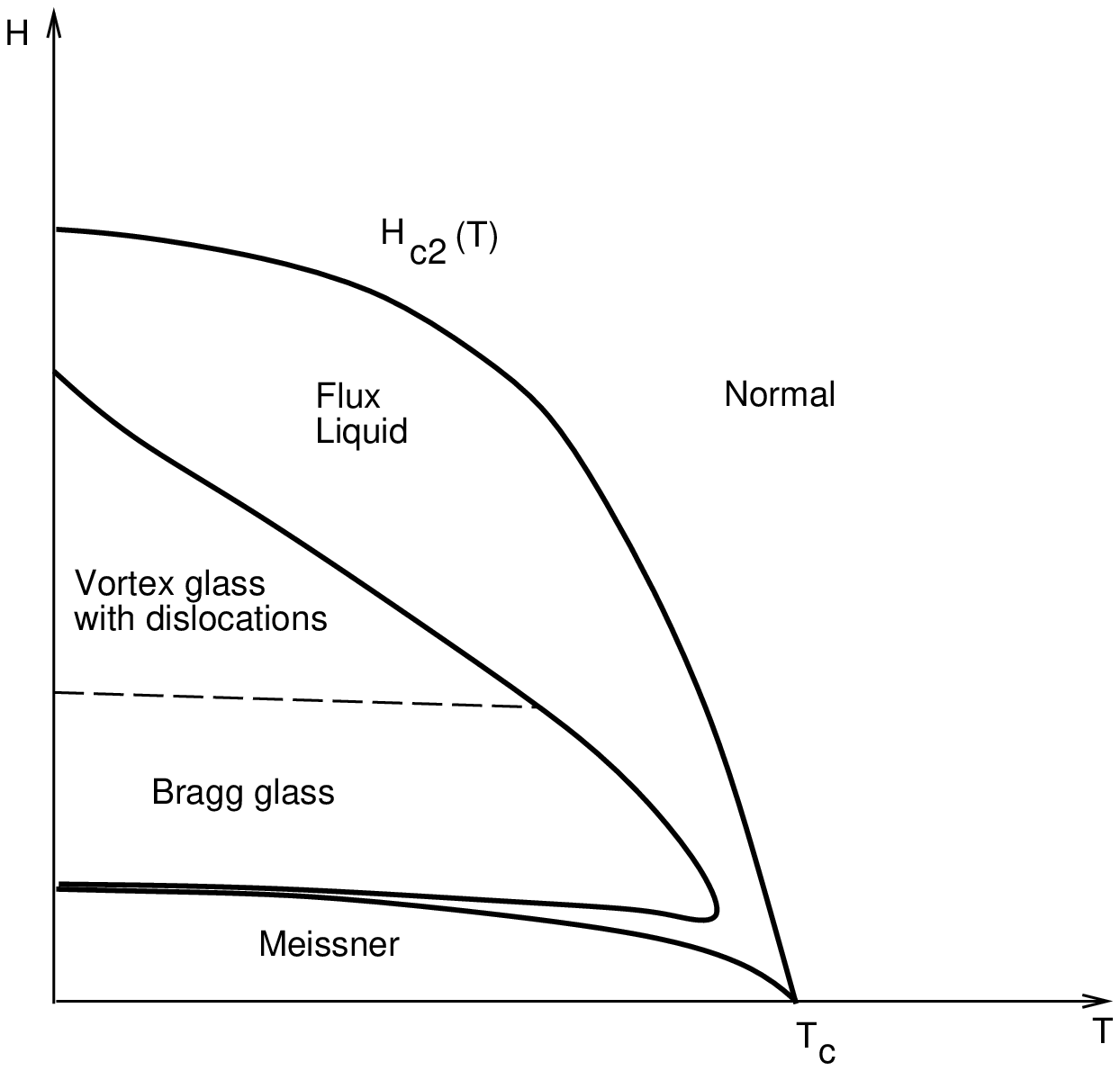}
\caption[]{
Phase diagram of the flux line system in high temperature superconductors.
Upper panel: in absence of disorder. Lower panel: in presence of disorder.
}\label{Fig_Abrikosov2}     
\end{figure}

In contrast to the usual phase diagram of conventional, low temperature superconductors, 
which shows an Abrikosov flux-line lattice on the whole 
region $H_{c1}(T)<H<H_{c2}(T)$ delimited by the lower and upper critical fields, $H_{c1}(T)$ and 
$H_{c2}(T)$ respectively (Fig. \ref{Fig_Abrikosov1}), because thermal fluctuations are now 
stronger, the vortex lattice melts into a flux-line liquid 
over a significant region of the magnetic field ($H$)-temperature ($T$) phase diagram (Fig.
\ref{Fig_Abrikosov2}, upper panel). 

In the presence of disorder, most theoretical studies of flux pinning in HTSCs have concentrated 
on trying to understand nature of translational and orientational order of flux-lines
as a function of temperature and applied external field, and the implications of the possible 
existence of glassy phases on the 
nonequilibrium dynamics, especially in the presence of an external current.
These studies also led to an investigation of these questions in the more general context of 
elastic media in the presence of thermal fluctuations, pinning (random and periodic) and external 
drive, stimulated by strong connections to other mathematically related condensed matter systems 
such as charge density waves (CDWs), Wigner crystals, domain walls, crystal surfaces, etc. 
For the particular case of flux-lines in presence of disorder, the phase diagram turns out to be 
a very rich one, with the
flux-line lattice losing true translational long range order (LRO) and exhibiting a highly
nonlinear flux flow resistivity. For weak disorder and low applied magnetic fields, this is a
phase with quasi long range order (QLRO) and no topological
defects
which has been termed Bragg glass. At higher external fields, the Bragg glass leads way to a
vortex glass with dislocations that improve the benefit from pinning energy
(Fig. \ref{Fig_Abrikosov2}, lower panel).

On the historical level, the first and decisive step for studying the pinning problem using 
statistical mechanics
was done in the remarkable work of Larkin\cite{Larkin,Larkin-Ovchinnikov}, where it was recognized that pinning can 
be treated perturbatively inside finite size domains (the Larkin domains) of coherently pinned flux lines, whose 
spatial extent has become known as the Larkin length $R_c$. Beyond this length, 
the disorder Hamiltonian has a large number of low lying metastable ``ground'' states,
which dooms to eventual failure all direct methods based on straightforward 
perturbation theory. Therefore, new, more sophisticated methods had to be invented to treat the
pinning problem beyond the Larkin length $R_c$.

The key quantity characterizing LRO in disordered elastic media is the roughness
\begin{eqnarray}
C({\bf r}) = \overline{\langle[{\bf u}({\bf r}) - {\bf u}({\bf 0})]^2\rangle} \,,
\label{def-CK}
\end{eqnarray}
where ${\bf u}({\bf r})$ is the displacement of a vortex from its undistorded position ${\bf r}$,
and where the angular brackets denote averaging over thermal fluctuations and 
the overbar denotes averaging over the quenched disorder.
Within a single Larkin correlated volume, one can neglect the dependence of the random pinning force 
${\bf f}({\bf u},{\bf r})$ experienced by flux lines on the displacements ${\bf u}$. 
Simple analysis\cite{Larkin} (see below) shows that $C(r)\sim r^{2\zeta}$ with the so-called 
wandering or roughness exponent $\zeta=(4-d)/2$ ($d$ here is the dimension of the lattice). Beyond the Larkin 
length $R_c$, we cannot neglect the dependence of ${\bf f}({\bf u},{\bf r})$ on the displacements ${\bf u}$, and the
periodicity of the flux line system has also to be taken into account.
Following the pioneering work of Larkin (and analogous works by Lee and Rice\cite{LeeRice} in the context of CDWs and 
by Imry and Ma\cite{ImryMa} in the context of random ferromagnets),
significant progress in understanding the physics on scales longer than $R_c$ was achieved thanks to 
the contributions of several authors, such as Villain and Fernandez\cite{Villain-Fernandez}, 
D.S. Fisher\cite{DSFisher}, Nattermann\cite{Nattermann}, Bouchaud, M\'ezard and Yedidia
\cite{Bouchaud-Mezard-Yedidia} and Korshunov\cite{Korshunov}. 
These results were then fruitfully applied to the flux line lattice problem
where it was shown by Giamarchi and Le Doussal\cite{Giamarchi1} 
(using a variant of the replica variational method for elastic manifolds of M\'ezard and 
Parisi\cite{Mezard})
that at length scales longer than 
$R_c$ the flux array is characterized by a logarithmic growth of flux line displacements
\begin{eqnarray}
C({\bf r}) \sim \ln(r/R_c)
,\label{logrough}
\end{eqnarray}
both in two and three dimensions, a result which has been confirmed by an independent, functional 
renormalization group (FRG) calculation.\cite{Emig}

An issue which is not sufficiently appreciated in the literature at the present time is the fact 
that the phonon field ${\bf u}({\bf r})={\bf u}({\bf x},z)$
includes all the internal modes of the flux lines, and that positional order is in fact best 
characterized by the displacement of the center of mass (CM) positions of vortices, which 
represent the average positions of the flux-lines. Indeed, flux-line lattices are essentially 
two-dimensional, as exhibited by the fact that both direct and reciprocal lattice vectors of the 
FLL lie in two-dimensional space. This is an {\em essential} difference between vortex lattices 
and ordinary three-dimensional solids, which implies in particular that in order to characterize the 
displacement of the average position of a given flux-line from its equilibrium position ${\bf 
R}_i$, it is enough to consider the displacement ${\bf u}_{0i}={\bf r}_{0i}-{\bf R}_i$
of the vortex center of mass ${\bf r}_{0i}$. Hence, to find the displacement of the average 
(i.e. center of mass) position of vortices, we need to consider the correlator
\begin{eqnarray}
C_0({\bf r}) = \overline{\langle[{\bf u}_0({\bf r})-{\bf u}_0({\bf 0})]^2\rangle} .
\end{eqnarray}
Although the above distinction between the general correlator $C({\bf r})$ 
of Eq. (\ref{def-CK}) and the CM correlator $C_0({\bf r})$ is of no great importance
and only of academic interest for pure (unpinned) flux line lattices in three 
dimensions, below we will show that proper handling of the CM mode of flux lines alters the 
asymptotic behavior of the translational correlation function (\ref{def-CK}) in the presence of 
disorder. In order to do so, we shall extend the replica Gaussian variational method of reference
\cite{Giamarchi1} to the case of samples of finite thickness $L_z$ along the direction of flux
lines, being careful to separate the CM mode from the internal modes of flux lines.

Our main result is that for the fluctuations of the center of mass positions of flux lines 
the logarithmic roughness of Eq. (\ref{logrough}) 
should be replaced, in finite size systems (and especially in thin films) by the following 
algebraic growth of flux-line fluctuations
(here the vortices are directed along the
$\hat{\bf z}$ axis, and ${\bf r}_\perp=x\hat{\bf x}+y\hat{\bf y}$)
\begin{eqnarray}
C_0(r_\perp) \sim \frac{r_\perp}{L_z} \,.
\label{algebraic}
\end{eqnarray}
Indeed, the main thrust of the argument made in ref. 13 relied on the fact that the elastic 
propagator $G(q)$ in presence of disorder had a wavevector dependence of the form
\begin{eqnarray}
G(q) \propto \frac{1}{q^d}
\label{G(q)}
\end{eqnarray}
in $d$ dimensions for $2<d<4$, hence the result (\ref{logrough}). It is important to realize that 
the elastic propagator in Eq. (\ref{G(q)}) is valid, strictly speaking, for 
three-dimensional solids (although it has been derived in the context of vortex lines).
If we recall that 
$q^2=q_\perp^2+q_z^2$ (with ${\bf q}_\perp$ and $q_z$ the wavevectors along the directions 
perpendicular and parallel to the vortices, respectively), then it is easy to see from the above 
result that the elastic propagator $G_0(q)$ for the CM ($q_z=0$) mode is given by:
\begin{eqnarray}
G_0(q_\perp) \propto \frac{1}{q_\perp^d} \,,
\end{eqnarray}
which directly leads to Eq. (\ref{algebraic}) in $d=3$ dimensions.
While we argue that the above result, Eq. (\ref{algebraic}), is in principle valid for
samples of arbitrary thickness, it is mostly relevant to thin superconducting films
with a small enough thickness $L_z$ compared to their size $L_\perp$ in transverse directions
(as long as $L_z\gg\xi$, where $\xi$
is the coherence length along the direction of flux lines, so that
flux lines have both CM and internal fluctuations).
In such films, the above equation can be of direct experimental relevance and will lead to the
complete destruction of translational long range order on length scales $r_\perp>L_z$.

The rest of this paper is organized as follows. In Sec. \ref{lro-elas}, we shall start our 
investigation by carefully defining the CM and internal degrees 
of freedom of the flux lines and the associated phonon fields. We shall 
then briefly discuss the case of disorder-free flux line lattices in $2+1$ and $1+1$ dimensions. 
In Sec. \ref{lro-LRO}, we shall consider the effect of an 
external pinning potential, both perturbatively, and using a replica Gaussian variational approach.
In Sec. \ref{lro-moving}, we briefly comment on positional order for the CM mode for moving 
flux line arrays in disorder, and in Sec. \ref{lro-conc}  we will present our conclusions.

\section{Elasticity in the center of mass representation}
\label{lro-elas}

\begin{figure}[h]
\includegraphics[scale=1.0]{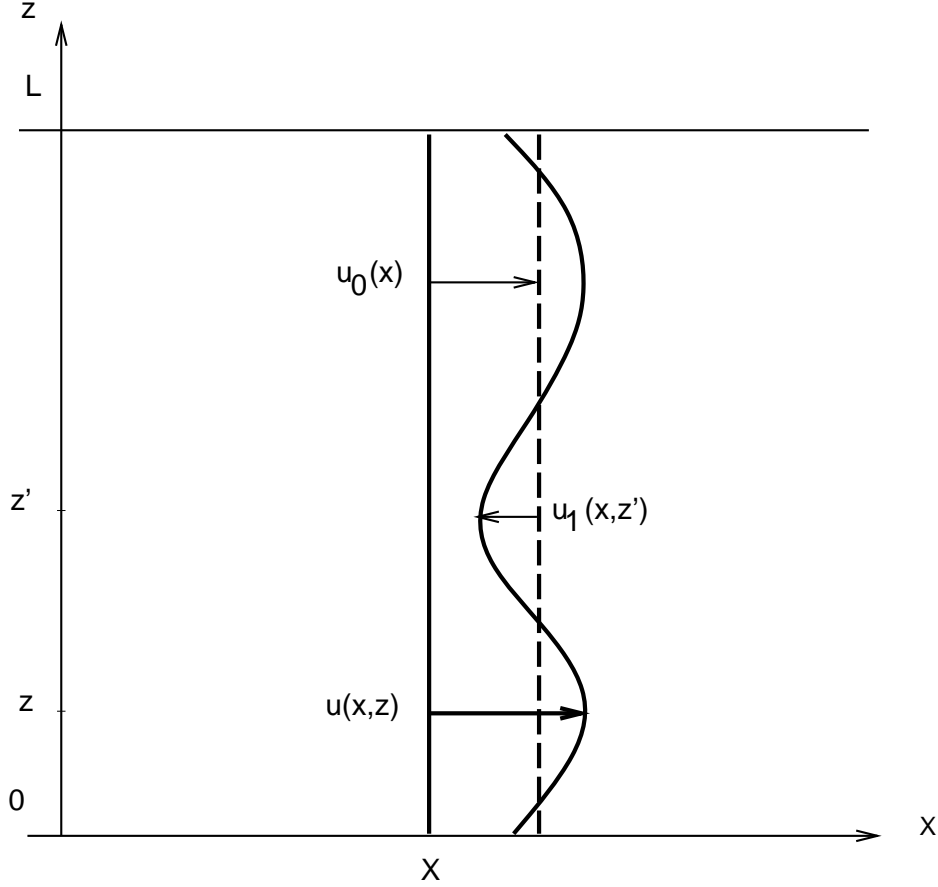}
\caption[Definition of the center of mass and internal modes of flux lines]{
Definition of the center of mass and internal modes. The solid and dashed vertical lines indicate the
location of the equilibrium and center of mass positions, respectively. ${\bf u}_0({\bf x})$ is the displacement
of the center of mass position with respect to the ideal position ${\bf x}$.
${\bf u}_1({\bf x},z)$ is the displacement of the flux line with respect to the 
center of mass position at height $z$, and ${\bf u}({\bf x},z)$ is the diplacement of the flux line 
at height $z$ with respect
to its equilibrium position. One has ${\bf u}({\bf x},z)={\bf u}_0({\bf x})+{\bf u}_1({\bf x},z)$.
}\label{Fig_CM}
\end{figure}

In order to fix ideas, let us consider a $d$-dimensional superconducting sample in an applied magnetic field 
${\bf H}=H\hat{\bf z}$. The sample thickness in the $z$ direction will be denoted by $L$.
Vortex trajectories will be parametrized by the $d$-dimensional vector
${\bf r}_i(z)=({\bf x}_i(z),z)$, where ${\bf x}_i(z)$ denotes the transverse position of the $i$th flux line
at height $z$, and the number of transverse dimensions will be denoted by $d_\perp=(d-1)$. 
We shall introduce the following decomposition of ${\bf x}_i(z)$ in Fourier 
modes\cite{Doi-Edwards,Radzihovsky-Frey}
\begin{eqnarray}
{\bf x}_i(z) = \sum_{n=-\infty}^\infty {\bf x}_i(q_n)\,\mbox{e}^{i{\bf q}_nz} \,,
\end{eqnarray}
where $q_n=2n\pi/L$ and 
where the Fourier components ${\bf x}(q_n)$ are related to ${\bf x}_i(z)$ by
\begin{eqnarray}
{\bf x}(q_n) = \frac{1}{L}\int_0^L dz\;{\bf x}_i(z)\,\mbox{e}^{iq_nz} \,,
\end{eqnarray}
as can be verified by using the orthogonality relation:
\begin{eqnarray}
\int_0^Ldz\;\mbox{e}^{iq_nz}\big(\mbox{e}^{iq_mz}\big)^*  =  L\,\delta_{n,m}\,.
\label{ortho-z}
\end{eqnarray}
The above Fourier decomposition is similar to the decomposition of internal modes into Rouse modes 
commonly used in polymer physics\cite{Doi-Edwards}.
For the developments that will follow, it will prove useful to write ${\bf x}_i(z)$ in the form
\begin{eqnarray}
{\bf x}_i(z) = {\bf x}_{0i} + {\bf u}_{1i}(z) \,,
\end{eqnarray}
where
\begin{eqnarray}
{\bf x}_{0i} = \frac{1}{L}\int_0^L dz\;{\bf x}_i(z)
\end{eqnarray}
is the position of the center of mass of the $i$th flux line, while (here ${c.c.}$ denotes complex conjugation)
\begin{eqnarray}
{\bf u}_{1i}(z) = \sum_{n=1}^\infty\{ {\bf x}(q_n)\,\mbox{e}^{iq_nz} + c.c.\}
\end{eqnarray}
is the displacement of the flux-line at height $z$ with respect to the CM position
${\bf x}_{0i}$. Note that this last quantity itself is a dynamical variable, since the CM position
${\bf x}_{0i}$ of the $i$-th flux line itself fluctuates around its ideal lattice position ${\bf X}_i$.    
We thus see that the displacement of the flux line at height $z$ with respect to its equilibrium position 
${\bf X}_i$ is given by
\begin{eqnarray}
{\bf u}_i(z) & = & {\bf x}_i(z) - {\bf X}_i \,, \nonumber\\
& = & {\bf u}_{0i} + {\bf u}_{1i}(z) \,,
\end{eqnarray}
where the $z$-independent quantity ${\bf u}_{0i}={\bf x}_{0i} - {\bf X}_i$ is the displacement 
of the center of mass position with respect to the equilibrium position ${\bf X}_i$ (see figure 
\ref{Fig_CM}).

In order to be able to define mathematically tractable models, 
one has to perform a continuum limit, whereby the individual displacement fields 
$\{ {\bf u}_i(z)\}$ are replaced by a smooth interpolating function ${\bf u}({\bf x},z)$. 
The displacement field ${\bf u}({\bf x},z)$ will be decomposed in the following Fourier 
modes:\cite{Giamarchi2}
\begin{eqnarray}
{\bf u}({\bf x},z) = \sum_{n=-\infty}^{\infty}\int_{\bf q} {\bf u}({\bf q},q_n)
\;\mbox{e}^{i({\bf q}\cdot{\bf x} + q_nz)}\,,
\end{eqnarray}
where $\int_{\bf q}$ stands for $\int_{BZ}d^{d_\perp}{\bf q}/(2\pi)^{d_\perp}$
(note that ${\bf q}$ here is a $d_\perp$-dimensional vector), and the integration is over the 
first Brillouin zone of the flux line lattice.
The Fourier components $\{ {\bf u}({\bf q},q_n)\}$ can be related to ${\bf u}({\bf x},z)$ through:
\begin{eqnarray}
{\bf u}({\bf q},q_n) = \frac{1}{L}\int_0^L dz \int\!d{\bf x}\;{\bf u}({\bf x},z)
\;\mbox{e}^{-i({\bf q}\cdot{\bf x} + q_nz)}\,,
\end{eqnarray}
as can be verified by using equation (\ref{ortho-z}) and the fact that 
\begin{eqnarray}
\int\!d{\bf x}\;\mbox{e}^{i({\bf q}+{\bf q}')\cdot{\bf x}}
= (2\pi)^{d_\perp}\,\delta({\bf q}+{\bf q}') \,.
\nonumber
\end{eqnarray}

In the above representation of the elastic displacements ${\bf u}({\bf x},z)$, the usual expression of the elastic 
matrix\cite{Brandt,Blatter-et-al} (here $P^L_{\alpha\beta}({\bf q})=q_\alpha\,q_\beta/q^2$ and
$P^T_{\alpha\beta}({\bf q})=\delta_{\alpha\beta}-P^L_{\alpha\beta}({\bf q})$ are the usual
projection operators)
\begin{eqnarray}
\Phi_{\alpha\beta}({\bf q},q_n) & = & \Phi_L({\bf q},q_n)P^L_{\alpha\beta}({\bf q}) +
\Phi_T({\bf q},q_n)P^T_{\alpha\beta}({\bf q})\,,
\end{eqnarray}
remains unchanged, except for the appearance of discrete $q_n$ modes (instead of a continuous $q_z$ variable) 
and for a slight change in the expression of the longitudinal and transverse components which, 
due to our specific Fourier transform convention, are now given by~:
\begin{eqnarray}
\Phi_L({\bf q},q_n) & = & L(c_{11}q^2 + c_{44}q_n^2) \,,\label{def-PhiL}
\\
\Phi_T({\bf q},q_n) & = & L(c_{66}q^2 + c_{44}q_n^2)  \,.\label{def-PhiT}
\end{eqnarray}
In the above definitions, $c_{11}$, $c_{66}$ and $c_{44}$ are the compression, shear and tilt moduli of the FLL, 
respectively. In principle, the compression and tilt moduli have spatial dispersion, i.e. they are 
wavevector-dependent\cite{Brandt}. Here we shall neglect this dispersion altogether, as it does not affect the 
long distance behavior of correlation functions.

The elastic Hamiltonian can now be written in the form
\begin{eqnarray}
H_{el} = \frac{1}{2}\sum_{n=-\infty}^\infty\int_{{\bf q}}\!\! u_{\alpha}(-{\bf q},-q_n)
\Phi_{\alpha\beta}({\bf q},q_n)u_{\beta}({\bf q},q_n) \,.
\label{elas2}
\end{eqnarray}
It is to be noted that
the elastic matrix $\Phi({\bf q},q_n=0)$ of the center of mass phonons in three dimensions
\begin{eqnarray}
\Phi_{\alpha\beta}({\bf q},0) = Lc_{11}\,q^2\;P^L_{\alpha\beta}({\bf q}) + 
Lc_{66}\,q^2\;P^T_{\alpha\beta}({\bf q})
\end{eqnarray}
has the same form as the elastic matrix of a two-dimensional lattice of ordinary particles with 
(two-dimensional) compression 
and tilt moduli given by $C_{11}=Lc_{11}$ and $C_{66}=Lc_{66}$ respectively.
From the expression (\ref{elas2}) of the elastic Hamiltonian, we can easily write for the thermal average 
$\langle u_{\alpha}({\bf q},q_n)u_{\beta}({\bf q}',q_m)\rangle$ the following result
(we henceforth use units such that Boltzmann's constant $k_B=1$)
\begin{eqnarray}
\langle u_{\alpha}({\bf q},q_n)u_{\beta}({\bf q}',q_m)\rangle & = & 
(2\pi)^{d_\perp}\delta_{n,-m}\delta_2({\bf q}+{\bf q}') 
\;TG_{\alpha\beta}({\bf q},q_n) \,,
\end{eqnarray}
with the elastic propagator
\begin{eqnarray}
G_{\alpha\beta}({\bf q},q_n) & = & [\Phi_L]^{-1}P^L_{\alpha\beta}({\bf q})
+[\Phi_T]^{-1}P^T_{\alpha\beta}({\bf q}) .
\label{def-Gel}
\end{eqnarray}
If we call $G_{0,\alpha\beta}({\bf q})$ the elastic propagator of the center of mass mode 
$G_{\alpha\beta}({\bf q},q_n=0)$, then 
it can easily be verified that the mean square relative displacement
$C({\bf x})=\big\langle[{\bf u}({\bf x})-{\bf u}({\bf 0})]^2\big\rangle$ 
can be written as the sum
\begin{eqnarray}
C({\bf x}) = C_0({\bf x}) + C_1({\bf x}) \,,
\end{eqnarray}
where
\begin{eqnarray}
C_0({\bf x})  =  2T\sum_\alpha\int_{\bf q} G_{0,\alpha\alpha}({\bf q})\big\{1-\cos[{\bf q}\cdot{\bf x}]\big\}
\label{BCM}
\end{eqnarray}
is the mean square displacement of the center of mass mode, while 
\begin{equation}
C_1({\bf x},z)\! = \! 2T\sum_\alpha\sum_{n\neq 0}\int_{\bf q} 
G_{\alpha\alpha}({\bf q},q_n)\big\{1-\cos[{\bf q}\cdot{\bf x}+q_nz]\big\}
\label{Binternal}
\end{equation}
is the corresponding quantity for the internal modes. In what follows, we shall be mainly interested in the center 
of mass relative displacement $C_0({\bf x})$ whose behavior has been largely overlooked in the past, and which 
will turn out to control the large scale asymptotics of $C({\bf x},z)$ and the long range translational order in 
the vortex system.

Using the properties of projection operators, one can easily verify that the 
trace of the elastic propagator ${\bf G}_{0}({\bf q})$ is given by~:
\begin{eqnarray}
\sum_{\alpha}G_{0,\alpha\alpha}({\bf q}) = \frac{1}{L}\Big(\frac{1}{c_{11}q^2} + \frac{1}{c_{66}q^2}\Big)\,,
\end{eqnarray}
and hence we obtain in $2+1$ dimensions~:
\begin{eqnarray}
C_0({\bf x}) \simeq  \frac{T}{{\pi}Lc_{66}}\int_0^\Lambda \frac{dq}{q}\;
\big[1-J_0(q|{\bf x}|)\big] \,,
\end{eqnarray} 
where we took into account the fact that $c_{66}\ll c_{11}$ (and hence that $c_{11}^{-1}$ is negligibly small 
compared to $c_{66}^{-1}$) for most HTSC.
The large distance behavior of the integral on the rhs of the last equation can be obtained in a   
standard way\cite{Chaikin-Lubensky}, with the result~:
\begin{eqnarray}
\int_0^\Lambda \frac{dq}{q}\;\big[1-J_0(q|{\bf x}|)\big] = \ln\Lambda|{\bf x}| + \kappa + 
{\cal O}\big((\Lambda|{\bf x}|)^{-3/2}\big) \,,\nonumber
\end{eqnarray} 
where 
$\kappa = \int_0^1 du\,(1-J_0(u))/u^2 \simeq -0.116 $
is a numerical constant.\cite{footnote1} 
Hence, the correlation function $C_0({\bf x})$ takes the form\cite{Nordborg-Blatter}
\begin{eqnarray}
C_0({\bf x})\simeq
\frac{T}{\pi Lc_{66}}\,\big[\ln\Lambda|{\bf x}| + \kappa\big] \,.
\end{eqnarray}
Using the fact that $c_{66}=\varepsilon_0/4a^2$, where $\varepsilon_0=(\phi_0/4\pi\lambda)^2$ 
(here $\phi_0=hc/2e$ is the flux quantum\cite{deGennes} and $\lambda$ is the London penetration depth) 
and $a$ is the vortex lattice spacing, 
we obtain
\begin{eqnarray}
\frac{1}{a^2}\;C_0({\bf x},z) \simeq
\Big(\frac{T}{T_0}\Big)\;\big[\ln\Lambda|{\bf x}| + \kappa\big] \,,
\label{qlro}
\end{eqnarray}
with the characteristic temperature
\begin{equation}
T_0 \simeq \frac{\pi}{4}\;L\varepsilon_0\,. \label{def-T0}
\end{equation}
The fact that $T_0$ is proportional to $L$ is an indication of the fact that the CM mode
with compression and shear moduli $C_{11}=Lc_{11}$ and $C_{66}=Lc_{66}$ 
respectively, is much stiffer than the internal modes. 
The characteristic temperature $T_0$ of equation (\ref{def-T0}) 
is in general a very large temperature for macroscopic sample thicknesses. 
Using the result\cite{Blatter-et-al}
\begin{equation}
\varepsilon_0\,(K/\AA)=\frac{1.964\times 10^8}{[\lambda(T)(\AA)]^2} 
= \frac{1.964\times 10^8}{[\lambda(0)(\AA)]^2}\,\big(1-\frac{T}{T_c}\big) \,,
\label{numerical-value-e0}
\end{equation}
with $\lambda(0)=1400\AA$ (typical of HTSC) and $L=1cm$, we find that $T_0$ is of the order of $10^{10}K$. Even in 
low temperature
superconductors with much higher values of $\lambda(0)$ (typically $\lambda(0)$ is of order $10^{4}$ to 
$10^{6}\AA$ in
these materials), $T_0$ is still very high for bulk samples. We therefore conclude that, even though it may appear at
first sight from equation (\ref{qlro}) that the center of mass positions of the vortices have only quasi long range
order, because the prefactor $(T/T_0)$ is extremely small for all temperatures of interest on one hand, and of the very
slow variation of the logarithm function on the other, it can be said that the center of mass mode possesses
true long range order for all realistic experimental situations.

The situation is, however, much less clear in $1+1$ dimensions (i.e. flux lines confined to a 
plane). There we find
\begin{eqnarray}
C_0(x) & = & \frac{T}{L}\int_{-\Lambda}^{\Lambda}\frac{dq}{2\pi}\,
\frac{1-\mbox{e}^{iqx}}{c_{11}q^2} \,,\nonumber\\
& = & \frac{T|x|}{\pi L c_{11}}\int_0^{\Lambda|x|}du\,\,\frac{1-\cos u}{u^2} \,,
\nonumber\\
& \simeq & \frac{T|x|}{2 L c_{11}} \,,
\label{B0-1+1}
\end{eqnarray}
where in going from the first to the second line we used the change of variables $q={\Lambda}|{\bf x}|$, 
and where we obtained the last line by sending $\Lambda|{\bf x}|$ to infinity and using 
the result\cite{Abramowitz}
$\int_0^\infty du(1-\cos u)/u^2=\pi/2$.
Here again, the survival of positional order in a given sample of thickness $L$ will depend on the numerical 
value of $(T/{\pi}Lc_{11})$.
If the vortices are so far apart from each other that the condition $a>\lambda$ is satisfied, then one can use the
following expression for the compression modulus\cite{Pokrovsky,Blatter-et-al}
\begin{eqnarray}
c_{11} = \frac{\pi^2 T^2n^4}{c_{44}}\,\frac{1}{(1-Dn)^2} \,,
\end{eqnarray}
where $n=1/a$ is the density of flux lines, and $D\approx\lambda$ is the range of the interaction between vortices. 
Assuming that the tilt modulus $c_{44}$ is of order $n\varepsilon_0$, and neglecting the factor $(1-Dn)^{-2}$ 
which is
of order unity for $a>\lambda$, we obtain\cite{Nattermann-Scheidl}
\begin{eqnarray}
c_{11} \simeq \frac{\pi^2T^2}{\varepsilon_0a^3} \,.
\end{eqnarray}
Using this expression in equation (\ref{B0-1+1}), we are led to the following numerical estimate
for the displacement correlation function:
\begin{eqnarray}
\frac{1}{a^2}C_0(x) = \frac{1.964\times 10^8(\mbox{K}\cdot\AA)}{2\pi^2T L}
\;\Big(\frac{a}{\lambda(T)}\Big)^2\,\,\frac{|x|}{a} \,,
\end{eqnarray}
where the expression (\ref{numerical-value-e0}) of $\varepsilon_0$ has been used. 
Thus we see that, even at very low temperatures (of order, say, a few kelvins) and large sample thicknesses 
($L\sim 1\mbox{cm}=10^8\AA$), long range order will be destroyed on 
a relatively short distance, of order a few lattice constants $a$. This is an important result in view of 
the fact that it is a generally accepted fact that line lattices in $1+1$ dimensions have quasi-long range 
translational order\cite{Blatter-et-al}. 
Here we have just shown that there might be physical regimes where the very notion of an 
ordered lattice in the $1+1$ geometry becomes questionable, even in the absence of disorder.

For higher vortex densities, no closed form for the elastic constants $c_{11}$ and $c_{44}$ in a 
film of finite thickness, satisfying the correct boundary conditions for magnetic fields and 
supercurrents at the surface of the 
film\cite{Abrikosov'64}, seems to exist in the literature. 
This case obviously deserves further attention but will not be considered in any detail here. 
It is however clear that, if $c_{11}$ becomes of order $n\varepsilon_0$ when $a\ll\lambda$, then the
center of mass correlation function will behave like $\sim(T/T_0)|x|a$, with $T_0$ given by equation 
(\ref{def-T0}), and long range order
will be effectively recovered again at all temperatures and realistic sample sizes.

\section{Positional order in the presence of a random pinning potential}
\label{lro-LRO}
 
We now turn our attention to the more interesting problem of the positional order of the FLL in the presence of
disorder. To start with, we shall first apply Larkin's argument to the center of mass mode of the
flux lines, before presenting a more careful replica analysis in Section \ref{lro-replicas}.

\subsection{Larkin's argument applied to the center of mass mode}

Following Larkin's original work\cite{Larkin,Larkin-Ovchinnikov}, we assume that the action of the random 
impurities on the FLL can be represented by a random pinning force
${\bf F}({\bf r})$ whose probability distribution is assumed to be Gaussian with zero mean and correlations
(we remind the reader that the overbar denotes an average over the disorder)
\begin{eqnarray}
\overline{F_\alpha({\bf x},z)F_\beta({\bf x}',z')} =
W\,\delta_{\alpha\beta}\,\delta({\bf x}-{\bf x}')\delta(z-z'),\label{randomforce}
\end{eqnarray}
or, in Fourier space, 
\begin{eqnarray}
\overline{ F_\alpha({\bf q},q_n)F_\beta({\bf q}',q_m) } =
LW\,\delta_{\alpha\beta}\,(2\pi)^2\delta({\bf q}+{\bf q}')\,\delta_{n,-m} \,.
\end{eqnarray}
The Hamiltonian of the flux line system is now given by:
\begin{eqnarray}
H = H_{el} - \int_0^Ldz\!\!\int d{\bf x}\;{\bf F}({\bf x},z)\cdot{\bf u}({\bf x},z)\,,
\label{Htotal}
\end{eqnarray}
where $H_{el}$ is the elastic Hamiltonian of equation (\ref{elas2}).
Note that, as it stands, the random pinning force ${\bf F}({\bf x},z)$ only depends
on the ``ideal'' positions of flux lines ``${\bf x}$'' and not on their displacements
${\bf u}({\bf x},z)$. As we will see below (in paragraph \ref{lro-replicas}), it is more appropriate
to represent the random impurities by a random pinning potential $V\big({\bf x}+{\bf u}({\bf x},z)\big)$,
and the random force term in equation (\ref{Htotal}) can be thought of as the leading term in a Taylor expansion of the 
random potential in terms of the phonon field $\{ {\bf u}({\bf x},z)\}$.

At low temperatures, the statistical mechanics associated with the Gibbs measure $\exp(-H/T)$ is dominated by 
those configurations of the displacement field $\{{\bf u}({\bf x},z)\}$ for which $H$ is 
minimal\cite{Larkin-Ovchinnikov,Brandt-corr,Chudnovsky}.
Minimization\cite{footnote2}
of $H$ with
respect to $\{{\bf u}({\bf q}, q_n)\}$ leads to the result
\begin{eqnarray}
u_\alpha({\bf q}, q_n) = G_{\alpha\beta}({\bf q},q_n)F_\beta({\bf q},q_n),\label{u-dis1}
\end{eqnarray}
where
$G_{\alpha\beta}({\bf q},q_n)$ is the elastic ptopagator of equation (\ref{def-Gel}).
It then follows that the correlation function $\langle u_\alpha({\bf r})u_\beta({\bf r}')\rangle$ 
averaged over the configurations of the random force (\ref{randomforce}) is given by:
\begin{eqnarray}
\overline{ \langle {u_\alpha}({\bf r})u_\beta({\bf r}')\rangle } & = & 
\sum_{n,m}\int_{\bf q}\int_{\bf q'}
G_{\alpha\gamma}({\bf q},q_n)G_{\beta\delta}({\bf q}',q_m) \times
\nonumber\\
&\times& \overline{ F_\alpha({\bf q},q_n)F_\beta({\bf q}',q_m) }
\mbox{e}^{i({\bf q}\cdot{\bf x}+q_nz)}\,\mbox{e}^{i({\bf q}\cdot{\bf x}'+q_mz')} \,.
\end{eqnarray}
Using expression (\ref{randomforce}) of the random force correlator, we obtain the following result
\begin{eqnarray}
\langle u_\alpha({\bf r})\,u_\beta({\bf r}')\rangle & = &
WL\sum_n\int_{\bf q} G_{\alpha\gamma}({\bf q},q_n)
G_{\beta\gamma}(-{\bf q},-q_n)
\mbox{e}^{i{\bf q}\cdot({\bf x}-{\bf x}')}\,\mbox{e}^{iq_n(z-z')}
,\label{uu-dis-larkin}
\end{eqnarray}
which implies that the relative displacement correlator 
$C({\bf x},z)=\langle[{\bf u}({\bf x},z)-{\bf u}({\bf 0},0)]^2\rangle$ is given by
\begin{eqnarray}
C({\bf x},z)& = & 2WL\sum_n\int_{\bf q}
G_{\alpha\beta}({\bf q},q_n)G_{\alpha\beta}(-{\bf q},-q_n)
\;[1-\mbox{e}^{i({\bf q}\cdot{\bf x}+q_nz)}] \,.
\end{eqnarray}
Now if we use expression (\ref{def-Gel}) of the elastic propagator $G({\bf q},q_n)$, we find that
the usual expression\cite{Larkin,Brandt-corr,Blatter-et-al} of the correlator $C({\bf x},z)$
in $d$ dimensions
\begin{eqnarray}
C({\bf x},z)\simeq 2W\int\frac{d^d{\bf k}}{(2\pi)^d}\,\,
\frac{1-\cos({\bf k}_\perp\cdot{\bf x}+k_zz)}{\big[c_{66}k_\perp^2 + c_{44}(k)k_z^2\,\big]^2}
\label{oldcorr}
\end{eqnarray}
is replaced by (we remind the reader that $d_\perp=d-1$)
\begin{eqnarray}
C({\bf x},z) \simeq \frac{2W}{L}\sum_{n=-\infty}^\infty
\int\frac{d^{d_\perp}{\bf q}}{(2\pi)^{d_\perp}}\,
\frac{1-\cos({\bf q}\cdot{\bf x}+q_nz)}{\big[c_{66}q^2 + c_{44}q_n^2\,\big]^2} \,.
\label{newcorr-larkin}
\end{eqnarray}
Obviously, if we transform the summation in this last expression into an integral we recover the 
$d$-dimensional result (\ref{oldcorr}), with the long distance behavior $C({\bf r})\sim r^{4-d}$.
However, here we are mainly interested in the fluctuations of the CM mode which, 
we would like to argue, are the most relevant ones for positional order in samples of finite thickness 
(more specifically, in samples whose thickness $L$ along the direction of flux lines is much smaller than
their transverse size $L_\perp$). 
We shall therefore write
\begin{eqnarray}
C({\bf x},z) = C_0({\bf x}) + C_1({\bf x},z) \,,
\end{eqnarray}
where 
\begin{eqnarray}
C_0({\bf x}) =
\frac{2W}{L}\int\frac{d^{d_\perp}{\bf q}}{(2\pi)^{d_\perp}}\,\,
\frac{1-\cos({\bf q}\cdot{\bf x})}{c_{66}^2q^4}
\end{eqnarray}
is the correlator of the center of mass positions, while 
\begin{eqnarray}
C_1({\bf x},z) = \frac{2W}{L}\sum_{n \neq 0}
\int\frac{d^{d_\perp}{\bf q}}{(2\pi)^{d_\perp}}\,\,
\frac{1-\cos({\bf q}\cdot{\bf x}+q_nz)}{\big[c_{66}q^2 + c_{44}q_n^2\,\big]^2}
\nonumber
\end{eqnarray}
corresponds to the internal modes of the FLL. Taking $d_\perp=2$ and 
transforming the above sum into an integral, we obtain the usual 
large distance result in $d=3$ dimensions\cite{Larkin,Brandt-corr,Blatter-et-al} 
(recall that here, we neglect the dispersion of the elastic 
constants; taking the dispersion of $c_{44}$ into account leads to a 
different behavior at short distances):
\begin{eqnarray}
C_1({\bf x},z)\simeq 
\frac{W\lambda}{2\pi^2 c_{66}^{3/2}c_{44}^{1/2}  }
\Big(\frac{|{\bf x}|^2}{\lambda^2} + \frac{a^2z^2}{\lambda^4}\Big)^{1/2}\,.
\label{corr-internal}
\end{eqnarray}
On the other hand, we have for the center of mass positions~:
\begin{eqnarray}
C_0({\bf x})& = & \frac{W}{\pi L c_{66}^2}\int_{1/L_\perp}^\Lambda dq
\;\frac{1-J_0(q|{\bf x}|)}{q^3} \,,
\nonumber\\
& = & \frac{W |{\bf x}|^2}{\pi L c_{66}^2}\int_{|{\bf x}|/L_\perp}^{\Lambda |{\bf x}|} du\,\,\frac{1-J_0(u)}{u^3}\,,
\label{cmcorr}
\end{eqnarray}
where, in going from the first to the second line we made the change of variables $u=qx$, and where we used 
the inverse of the transverse size of the system, $1/L_\perp$, as an infrared cut-off to insure the convergence of 
the integral as $q\to 0$. 
At large distances, the upper bound of the integral on the rhs of equation (\ref{cmcorr}) can be
replaced by infinity, with the result (we assume that $|{\bf x}|\ll L_\perp$):
\begin{eqnarray}
\int_{|{\bf x}|/L_\perp}^{\Lambda|{\bf x}|} du\,\,\frac{1-J_0(u)}{u^3} \simeq 
\int_{|{\bf x}|/L_\perp}^1 du\,\,\frac{1-J_0(u)}{u^3} + 
\int_1^{\infty} du\,\,\frac{1-J_0(u)}{u^3} \,.
\label{int1-cmcorr}
\end{eqnarray} 
Now, on the interval $[x/L_\perp,1]$, we can approximate $J_0(u)=1-u^2/4+\mbox{o}(u^4)$. This leads to
\begin{eqnarray}
\int_{x/L_\perp}^1 du\,\,\frac{1-J_0(u)}{u^3} \approx \frac{1}{4}\,\ln\big({L_\perp}/{x}\big)\,.
\end{eqnarray}
On the other hand, $\int_1^{\infty} du\,\,\frac{1-J_0(u)}{u^3}$ is just a numerical constant, whose value we
shall denote by $\alpha$, and which is approximately given by $\alpha\simeq 0.287$, so that equation (\ref{cmcorr})
finally yields~:
\begin{eqnarray}
C_0({\bf x})
\simeq \frac{W |{\bf x}|^2}{4\pi L c_{66}^2}
\Big[\,\ln\big({L_\perp}/{x}\big)+\alpha'\,\Big] ,\label{resultcmcorr}
\end{eqnarray}
where we defined $\alpha'=4\alpha\simeq 1.15$. Apart from the unimportant numerical constant
$\alpha'$ (which can be absorbed in the cut-off $R$), this result is very similar to what has been obtained a long time
ago\cite{Brandt-corr,Kes} for thin superconducting films. Here we see that this result in fact holds true for the 
CM mode of the flux lines in samples of arbitrary thickness.

The above perturbative analysis is correct up to the length scale $x_c$ such that
$C_0(x_c)\simeq \xi^2$.
The quantity $x_c$ defines a length at which the center of mass mode of vortices has 
``random-walked'' a distance $\xi$ from a given initial position. Using equation 
(\ref{resultcmcorr}), we obtain that $x_c$ is the solution of 
the following equation (henceforth we omit the constant $\alpha$ from our considerations)
\begin{eqnarray}
x_c^2\simeq \frac{4\pi L c_{66}^2\xi^2}{W\ln(L_\perp/x_c)} \,.
\end{eqnarray}
A good approximation to this last quantity is obtained by replacing $x_c$ inside the argument of the 
logarithm by
$x_0=(4\pi L c_{66}^2\xi^2/W)^{1/2}$. We then have, to logarithmic accuracy:
\begin{eqnarray}
x_c \simeq \Big[\frac{4\pi L c_{66}^2}{W\ln(L_\perp/x_0)}\Big]^{1/2}\,\xi \,.
\end{eqnarray}
The important thing to note about $x_c$ is that it varies with the sample thickness $L$ as $L^{1/2}$, and hence is
very large for bulk samples. Let us for example compare $x_c$ to the in-plane characteristic length $R_c^\perp$ 
for the internal modes. For the sake of argument, we define $R_c^\perp$ as the length scale at which the equal 
height correlator
$C_1({\bf x},0)=W |{\bf x}|/(2\pi^2 c_{66}^{3/2}c_{44}^{1/2})$ reaches
the value $\xi^2$. We have
\begin{equation}
R_c^\perp = \frac{2\pi^2}{W}\;c_{66}^{3/2}c_{44}^{1/2}\,\xi^2 \,,
\end{equation}
and hence
\begin{eqnarray}
\frac{x_c}{R_c^\perp} &=& \frac{(L W)^{1/2}}{\pi^{3/2}  
\xi\,c_{66}^{1/2}c_{44}^{1/2}\;\sqrt{\ln(L_\perp/x_0)}} \,,
\nonumber\\
& = & \frac{(L W)^{1/2}\; a^3}{\pi^2\xi\lambda\varepsilon_0\sqrt{\ln(L_\perp/x_0)}} \,,
\end{eqnarray}
where, in going from the first to the second line, we used the fact that
$c_{44}=B^2/4\pi$ (with $B=\phi_0/a^2$ the magnetic induction inside the superconductor) and hence 
that\cite{Blatter-et-al} 
\begin{equation}
\sqrt{c_{66}c_{44}} = \frac{\pi\lambda\varepsilon_0}{a^3} \,.
\end{equation}
Now, if we use $W\approx \varepsilon_0^2/d_0^3$, where we denote 
by $d_0$ the average distance between impurities 
in the superconducting sample, we obtain
\begin{eqnarray}
\frac{x_c}{R_c^\perp} \approx 
\frac{\kappa}{\pi^2\sqrt{\ln(L_\perp/x_0)}}\,\,\Big(\frac{a}{\lambda}\Big)^2\,
\Big(\frac{a}{d_0}\Big)\,\sqrt{\frac{L}{d_0}} \,,
\end{eqnarray}
with $\kappa=\lambda/\xi$. Using $\kappa=100$, $\lambda(0)=1400\AA$, $a=500\AA$, $d=100\AA$ and $L_\perp=L=1\mbox{cm}$
we obtain
\begin{equation}
\frac{x_c}{R_c^\perp} \approx 6\times 10^2\,(1-T/T_c)\,.
\end{equation}
This shows that, now matter how large the internal fluctuations of the flux lines are, the 
center of mass mode of the flux lines experiences much smaller fluctuations and has a much 
larger Larkin length than the internal modes.

It is interesting to see how the above results are modified if correlated disorder is considered  
instead of point disorder. In that case, the random force correlations are given by:
\begin{eqnarray}
\overline{F_\alpha({\bf x})\,F_\beta({\bf x}')}
= W_c\;\delta_{\alpha\beta}\delta({\bf x}-{\bf x}') \,,
\end{eqnarray}
with\cite{Nelson-Vinokur} $W_c\approx \varepsilon_0^2/(a^2d_0^2)$. 
An immediate consequence of the fact that the random force does not depend on $z$ 
is that $F_\alpha({\bf x})$ will couple only to the center of mass mode of the flux lines, as can be easily seen 
using the fact that $\int_0^Ldz\;{\bf u}_1(z)=0$ in the last term of equation (\ref{Htotal}) (where we now let 
${\bf F}({\bf x},z)={\bf F}({\bf x})$).
The Hamiltonian of the flux line system can now be written in the form (here and in what follows we concentrate 
on $d=3$ dimensions)~:
\begin{eqnarray}
H & = & \frac{1}{2}\sum_{n\neq 0}\int_{{\bf q}}\!\! u_{\alpha}(-{\bf q},-q_n)
\Phi_{\alpha\beta}({\bf q},q_n)u_{\beta}({\bf q},q_n) + \nonumber\\
&+& \frac{1}{2}\sum_{n\neq 0}\int_{{\bf q}}\!\! u_{0,\alpha}(-{\bf q})
\Phi_{0,\alpha\beta}({\bf q})u_{0,\beta}({\bf q})
- L\int {d\bf x}\; {\bf F}({\bf x})\cdot{\bf u}_0({\bf x})
,\label{Ham-corr}
\end{eqnarray}
from which we obtain the interesting result (within the Larkin approximation of ignoring the ${\bf u}$ dependence
of the random pinning force ${\bf F}$)
that correlation functions for the internal modes fluctuations
remain unchanged and are in fact the same as their pure counterparts. Only the CM mode is affected by correlated 
disorder in first order perturbation theory, and the CM part of the Hamiltonian (\ref{Ham-corr}) 
maps onto the Hamiltonian of a two-dimensional system with compression and bulk moduli given by 
$C_{11}=Lc_{11}$ and $C_{66}=Lc_{66}$ respectively, subject to the random force 
$\tilde{\bf F}({\bf x}) = L {\bf F}({\bf x})$ whose variance is now given by $\tilde{W}_c=L^2W_c$. 
The relative displacement of the center of mass mode $C_0({\bf x})$ is then obtained in a very natural way as
\begin{eqnarray}
C_0({\bf x}) &=& 2\tilde{W}_c\,\int\frac{d^2\bf q}{(2\pi)^2}\,\,
G_{\alpha\beta}({\bf q},0)G_{\alpha\beta}(-{\bf q},0)
\;[1-\mbox{e}^{i{\bf q}\cdot{\bf x}}] \,,
\nonumber\\
& = & \frac{2W_c}{c_{66}^2}\,\int\frac{d^2\bf q}{(2\pi)^2}\,
\frac{1-\cos({\bf q}\cdot{\bf x})}{q^4}\,.
\end{eqnarray}
A calculation similar to the one carried out in equations 
(\ref{int1-cmcorr})-(\ref{resultcmcorr}) leads to the 
result
\begin{eqnarray}
C_0({\bf x}) = \frac{W_c|{\bf x}|^2}{4\pi c_{66}^2}\,\ln(L_\perp/x)
,\label{corr-columns}
\end{eqnarray}
which is the result obtained by Nelson and Vinokur\cite{Nelson-Vinokur}. Unlike the displacement correlator 
in the presence of point disorder, equation (\ref{resultcmcorr}), 
the correlator (\ref{corr-columns}) does not depend on the sample
thickness $L$. The resulting size $x_c$ of the perturbative region for the CM mode, which verifies the implicit 
equation
\begin{equation}
x_c = \frac{4\pi c_{66}^2\xi^2}{W_c\ln(L_\perp/x_c)}\,,
\end{equation}
does not increase with $L$, which translates the fact that columnar pins are much more effective in disrupting the
long range order of the FLL than point-like pinners.

\subsection{Variational analysis}
\label{lro-replicas}

As is well known, the Larkin analysis of the previous section breaks down beyond the Larkin 
length $x_c$. To find the long distance behavior of correlation functions, one has to resort to a 
more careful kind of analysis. Here we shall use the Gaussian variational 
method \cite{Mezard,Bouchaud-Mezard-Yedidia,Giamarchi1,Giamarchi2} (GVM)
to find the long distance behavior of the 
center of mass relative displacement $C_0({\bf x})$. 
The GVM has been used in the past to study the effect of disorder on elastic manifolds\cite{Mezard} and flux lattices
\cite{Bouchaud-Mezard-Yedidia,Giamarchi1,Giamarchi2}, and although uncontrolled (due to the absence of a small expansion parameter),
it has been shown\cite{Giamarchi1,Emig} to yield results that are in qualitative agreement with functional 
renormalization group calculations.

Here, we shall restrict ourselves to the three dimensional case and 
to isotropic elasticity, the generalization to the more realistic case of different elastic moduli being 
relatively straightforward. Following ref.\cite{Giamarchi1}, in this subsection we shall use the following Hamiltonian
\begin{eqnarray}
H  =  \int d{\bf x}dz\;\frac{1}{2}c\big[(\partial_z{\bf u})^2 + (\nabla_{\bf x}\cdot{\bf u})^2\big] + 
\int d{\bf x}dz\,\, V({\bf x},z)\rho({\bf x},z) ,\label{hamilt1}
\end{eqnarray}
where the vortex density is given by 
\begin{eqnarray}
\rho({\bf x},z) = \sum_i\delta_2({\bf x}-{\bf x}_i(z)) = 
\sum_i\delta_2\big({\bf x}-{\bf R}_i - {\bf u}_i(z)\big) \,,
\end{eqnarray}
and where $V({\bf x},z)$ is a Gaussian random potential with zero mean and correlations 
\begin{eqnarray}
\overline{ V({\bf x},z)V({\bf x}',z') } = \Delta({\bf x}-{\bf x}',z-z') \,.
\end{eqnarray}
Using the replica ``trick''
\begin{equation}
\overline{ \ln Z} = \lim_{n\to 0}\,\frac{\overline{Z^n} - 1}{n} \label{trick}
\end{equation}
to average the free energy over disorder, we obtain the following effective Hamiltonian 
\begin{eqnarray}
H_{eff} & = & \frac{1}{2}\sum_{a=1}^n \int d{\bf x}dz\,\,c\big[(\partial_z{\bf u}_a)^2 + 
(\nabla_{\bf x}{\bf u}_a)^2\big] + \nonumber \\
&-& \frac{1}{2T}\sum_{a,b}\int d{\bf x}dz\int d{\bf x}'dz'\;\rho_a({\bf x},z)\rho_b({\bf x}',z') 
\Delta({\bf x}-{\bf x}',z-z')\,.
\label{Heff-1}
\end{eqnarray}
Now, using the following decomposition of the density\cite{Haldane,Fisher,Giamarchi1}
\begin{eqnarray}
\rho({\bf x},z) \simeq \rho_0\,\big[1 - \partial_\alpha u_\alpha({\bf x},z) \big] + 
\rho_0\sum_{{\bf K}\neq 0}\rho_{\bf K}({\bf x})\mbox{e}^{i{\bf K}\cdot{\bf x}} \,,
\end{eqnarray}
where $\rho_{\bf K}({\bf x}) = \mbox{e}^{-i{\bf K}\cdot{\bf u}({\bf x},z)}$, and discarding rapidly oscillating
terms, we obtain from equation (\ref{Heff-1})~:
\begin{eqnarray}
H_{eff} & = & \frac{1}{2}\sum_{a=1}^n \int d{\bf x}dz\,\,c\big[(\partial_z{\bf u}_a)^2 +
(\nabla_{\bf x}{\bf u}_a)^2\big] 
\! -\! \sum_{a,b} \int d{\bf x} dz \,\Big\{
\frac{\Delta_0}{2T}\partial_\alpha u_\alpha^a({\bf x},z)\partial_\beta u_\beta^b({\bf x},z)
\nonumber\\
& + &\sum_{{\bf K}\neq 0} \frac{\Delta_{\bf K}}{2T}\,\cos\big({\bf K}\cdot
({\bf u}_a({\bf x},z)-{\bf u}_b({\bf x},z))\big)
\Big\} \,,
\end{eqnarray}
where we defined 
\begin{eqnarray}
\Delta_0 & = & \rho_0^2\int d{\bf x} \Delta({\bf x}) \,,
\\
\Delta_{\bf K} & = & \rho_0^2\int d{\bf x} \Delta({\bf x})\,\mbox{e}^{i{\bf K}\cdot{\bf x}} \,.
\end{eqnarray}
The highly nonlinear form of $H_{eff}$ precludes exact analysis. In order to make progress a common way is to use the 
so-called Gaussian variational method (GVM), which has been developed by M\'ezard and Parisi\cite{Mezard} in the 
context of random manifolds, and was first applied to the vortex lattice by Bouchaud, M\'ezard and Yedidia\cite{Bouchaud-Mezard-Yedidia}. 
This method consists in trying to find the best quadratic 
variational Hamiltonian $H_v$ to describe the full nonlinear problem using Boguliubov's
variational free energy\cite{Feynman-statmech,Chaikin-Lubensky}
\begin{eqnarray}
F_v = \langle H-H_v\rangle_v - T\ln Z_v \,,
\end{eqnarray}
where $\langle \cdots \rangle_v$ denotes averaging with statistical weight $\exp(-H_v/T)/Z_v$, and where 
$Z_v=\mbox{Tr}(\mbox{e}^{-H_v/T})$. As a trial Hamiltonian we take
\begin{eqnarray}
H_v = \frac{1}{2}\sum_n\int\frac{d^d{\bf q}}{(2\pi)^d} \,\,(G^{-1})_{ab}({\bf q},q_n){\bf u}_a({\bf q},q_n)
\cdot{\bf u}_b(-{\bf q},-q_n)
\end{eqnarray}
where $(G^{-1})_{ab}$ is an $n\times n$ matrix of variational parameters. The variational free energy is then
given by
\begin{eqnarray}
F_v & = &\frac{1}{2}\sum_{a,b}\sum_m\int_{\bf q}\big\{\big[c(q^2+q_m^2)\delta_{ab} -
\frac{\Delta_0}{T}\,q^2\big]\,G_{ab}({\bf q},q_m) - d_{\perp}T[\ln(TG)]_{aa}\delta_{ab} \big\}
\nonumber\\
& - &\sum_{a,b}\sum_{{\bf K}\neq 0} \frac{\Delta_K}{2T}\,\mbox{e}^{-\frac{1}{2}K^2C_{ab}({\bf x}=0,z=0)}\,,
\label{Fvar}
\end{eqnarray}
where we defined the difference correlation function (note that in this section, difference 
correlation functions $C({\bf x},z)$ are defined with an extra factor $1/d_\perp$, and that 
throughout the rest of this paper, no summation is 
implied on repeated indices)~:
\begin{eqnarray}
C_{ab}({\bf x},z) & = & \frac{1}{d_\perp}\;\langle[{\bf u}_a({\bf x},z)-{\bf u}_b({\bf 0},0)]^2\rangle
\nonumber\\
& = & T\sum_n\int_{\bf q} \big[
G_{aa}({\bf q},q_n) + G_{bb}({\bf q},q_n) -2\cos({\bf q}\cdot{\bf x}+q_nz)\,G_{ab}({\bf q},q_n)
\big].
\end{eqnarray}
Minimization of the variational free energy (\ref{Fvar}) with respect to $[G({\bf q},q_n)]_{ab}$ for $a\neq b$ 
leads to the following expression\cite{Giamarchi1}
\begin{eqnarray}            
[G^{-1}({\bf q},q_n)]_{ab} & = & c(q^2+q_n^2)\,\delta_{ab}
-\frac{\Delta_0}{d_{\perp}T} q^2 
\nonumber\\
& - & \sum_{K\neq 0}\frac{K^2\Delta_{\bf K}}{d_{\perp}T}
\exp\big(-\frac{1}{2}K^2C_{ab}({\bf x}=0,z=0)\big)\,.
\end{eqnarray}
Defining the self energy matrix $\sigma_{ab}$ by 
\begin{eqnarray}
[G^{-1}({\bf q},q_n)]_{ab} = c(q^2+q_n^2)\delta_{ab} - \sigma_{ab}\,,
\end{eqnarray}
we obtain
\begin{eqnarray}            
\sigma_{a\neq b} = +\frac{\Delta_0}{d_{\perp}T} q^2 +\sum_{K\neq 0}\frac{K^2\Delta_{\bf K}}{d_{\perp}T}
\exp\big(-\frac{1}{2}K^2C_{ab}({\bf x}=0,z=0)\big) \,,
\label{var-1}
\end{eqnarray}
while for $a=b$, we obtain
\begin{eqnarray}            
\sigma_{aa} = - \sum_{b\neq a}\sigma_{ab} \,.
\label{var-2}
\end{eqnarray}
Two types of solution can exist for the variational equations (\ref{var-1})-(\ref{var-2}). The first type is
the so-called ``replica-symmetric'' solution which preserves the permutation symmetry between replicas, while the 
second type is a ``replica-broken'' solution in which permutation symmetry between replicas is violated. 
Here we shall not give much details on the replica symmetric 
solution, and refer the reader to reference\cite{Giamarchi1} where more information can be found. 
The only change with respect to the latter reference is in the form of the diagonal correlation function 
$G_{aa}({\bf q},q_n)$ which, because the $q_z$ modes in our description are discrete, now takes the form
\begin{eqnarray}
G_{aa}({\bf q},q_n)  = \frac{1}{c(q^2+q_n^2)} + 
\frac{1}{c^2(q^2+q_n^2)^2d_{\perp}T}
\sum_{\bf K}K^2\Delta_{\bf K}\mbox{e}^{-K^2\ell^2/2} \,.
\label{diag-G}
\end{eqnarray}
In the above expression, $\ell$ is the (thermal) Lindemann length\cite{Giamarchi1}
\begin{eqnarray}
\ell^2 & = & \frac{2T}{L}\sum_{n=-\infty}^\infty\int_0^\Lambda\frac{d^{d-1}\bf q}{(2\pi)^{d-1}}
\frac{1}{c(q^2+q_n^2)} \,,
\nonumber\\
& \simeq & \frac{T}{\pi ca} \,, 
\end{eqnarray}
where the second line refers to $d=2+1$ dimensions. Due to the factor $1/(q^2+q_n^2)^2$ on the rhs of 
equation (\ref{diag-G}), we find that the center of mass relative displacements grow as
\begin{eqnarray}
C_{0,aa}({\bf x}) \sim x^{5-d} \,,
\end{eqnarray}
which, for $d=3$ is nothing but the Larkin result of eq. (\ref{resultcmcorr}).

The analysis of quadratic perturbations, in replica space, about the replica-symmetric solution described above 
is most conveniently done using the eigenvalue of the so-called ``replicon'' mode\cite{deAlmeida,Mezard}, and such 
an analysis (see Appendix \ref{lro-appendix}) shows that, for the particular case $d=3$, there 
exists 
a temperature $T_c=4\pi c L/K_0^2$
between a high-temperature, replica symmetric phase, and a low temperature glassy phase where replica symmetry is 
broken. This transition temperature is exactly the same as the one obtained previously in the context of 
correlated-disorder\cite{Giamarchi2}. Of course, for macroscopic samples $(L\gg a)$ this transition 
temperature $T_c$ is very large and is certainly not experimentally accessible, but from a conceptual point
of view it is important to realize its existence for three dimensional flux lattices in the presence of point disorder.

Now, in the low temperature, replica broken phase, the $n\to 0$ limit of matrices in replica space become
functions\cite{Parisi,Mezard} of a real variable $v$ which parametrizes pairs of low lying metastable 
states.\cite{footnote3}
Following\cite{Mezard,Giamarchi1} we let $\tilde G({\bf q},q_n)= G_{aa}({\bf q},q_n)$, 
$\tilde C({\bf x},z)=C_{aa}({\bf x},z)$, and parametrize $G_{ab}({\bf q},q_n)$ and 
$C_{ab}({\bf x},z)$ for $a\neq b$ by $G({\bf q},q_n;v)$ and $C({\bf x},z;v)$ respectively, with $0<v<1$. The 
saddle point equation 
(\ref{var-1}) can then be rewritten in the form~:
\begin{eqnarray}
\sigma(v) = \sum_{\bf K} \frac{\Delta_{\bf K}}{d_{\perp}T}\,K^2\mbox{e}^{-\frac{1}{2}K^2C(0,v)},\label{exp-sigma}
\end{eqnarray}
where we neglected the subdominant\cite{Giamarchi1} $K=0$ component of the disorder (the 
$\Delta_0q^2/d_{\perp}T$ term) and where
\begin{eqnarray}
C({\bf 0},0;v) \!=\! \frac{2T}{L}\sum_{n=-\infty}^\infty\int\frac{d^{d-1}\bf q}{(2\pi)^{d-1}}
\big(\tilde G({\bf q},q_n) - G({\bf q},q_n;v)\big) \,.
\nonumber
\end{eqnarray}
The algebraic rules for inversion of hierarchical matrices\cite{Mezard,Giamarchi1} give us~:
\begin{eqnarray}
C({\bf 0},0;v) = C({\bf 0},0;v_c) + 
\frac{1}{L}\sum_{n=-\infty}^\infty\int_v^{v_c} du\,\int\frac{d^{d-1}\bf q}{(2\pi)^{d-1}}
\,\,\frac{2T\sigma'(u)}{\big( c(q^2+q_n^2) + [\sigma](u)\big)^2}
\end{eqnarray}
where we defined $[\sigma](v) = v\sigma(v) - \int_0^vdu\,\sigma(u)$ and
\begin{eqnarray}
C({\bf 0},0;v_c) = \frac{1}{L}\sum_{n=-\infty}^\infty\int\frac{d^{d-1}\bf q}{(2\pi)^{d-1}}
\frac{2T}{ c(q^2+q_n^2) + [\sigma](v_c)}
\end{eqnarray}
Taking the derivative of equation (\ref{exp-sigma}) with respect to $v$ (keeping only\cite{Giamarchi1} the 
reciprocal lattice vectors ${\bf K}$ such that $K=K_0$), we obtain
\begin{eqnarray}
1 & = & \sigma(v)\cdot \frac{1}{L}\sum_{n=-\infty}^\infty \int\frac{d^{d-1}\bf q}{(2\pi)^{d-1}}\,\,
\frac{TK_0^2}{\big(c(q^2+q_n^2) + [\sigma](v)\big)^2} 
,\label{base_eqn}\\
& = & \frac{\sigma(v)}{L}\int\frac{d^{d-1}\bf q}{(2\pi)^{d-1}}\,\,
\frac{TK_0^2}{\big(cq^2 + [\sigma](v)\big)^2} 
+ \frac{\sigma(v)}{L}\sum_{n\neq 0}\,\, \int\frac{d^{d-1}\bf q}{(2\pi)^{d-1}}\,\,
\frac{TK_0^2}{\big(c(q^2+q_n^2) + [\sigma](v)\big)^2} \,.
\end{eqnarray}
Transforming the sum on the rhs into an integral, we obtain
\begin{eqnarray}
1 &\simeq& \frac{\sigma(v)}{L}\int\frac{d^{d-1}\bf q}{(2\pi)^{d-1}}\,\,
\frac{TK_0^2}{\big(cq^2 + [\sigma](v)\big)^2}
+  \sigma(v)\int\frac{d^d\bf k}{(2\pi)^d}\,\,
\frac{TK_0^2}{\big(ck^2 + [\sigma](v)\big)^2} \nonumber\\
& = & \frac{\sigma(v)}{L}\Big(\frac{TK_0^2 c_{d-1}}{ c^{(d-1)/2} }\Big)\,[\sigma(v)]^{(d-5)/2}
+  \sigma(v) \Big(\frac{TK_0^2 c_d}{c^{ d/2 }}\Big)\,[\sigma(v)]^{(d-4)/2} \,,\label{eq-sigma}
\end{eqnarray}
where the ultraviolet cut-off $\Lambda=2\pi/a$ has been sent to infinity,
and where we defined~:
\begin{equation}
c_d = \int\frac{d^d\bf k}{(2\pi)^d}\,\frac{1}{(k^2+1)^2}=\frac{(2-d)\pi^{1-d/2}}{2^{d+1}\sin(\pi d/2)\Gamma(d/2)}
\nonumber
\end{equation}
In the limit $L\to \infty$, the first term on the rhs of equation
(\ref{eq-sigma}) will
be negligibly small compared to the second one, and we can write
\begin{eqnarray}
1 \simeq \sigma(v) \Big(\frac{TK_0^2 c_d}{c^{d/2}}\Big)\,[\sigma(v)]^{(d-4)/2} \,.
\label{approx-sigma}
\end{eqnarray}
This approximation will turn out to be valid for a large range of values of the parameter $v$, covering almost the
entire interval $[0,1]$ when $L\to\infty$. Taking the derivative of equation (\ref{approx-sigma}) one more time with
respect to $v$, and using the fact that $[\sigma]'(v)= v\sigma'(v)$, we finally obtain\cite{Giamarchi1}
\begin{eqnarray}
[\sigma](v)  = \Big(\frac{v}{v_0}\Big)^{2/\theta} \,,
\label{solution-sigma}
\end{eqnarray}
where $\theta=d-2$ and $v_0 = 2TK_0^2c_d/((4-d)c^{d/2})$. Replacing the solution (\ref{solution-sigma}) into equation
(\ref{eq-sigma}),
one can verify that the first term on the rhs of that last equation can indeed be neglected if
\begin{eqnarray}
(v_0/v)^{1/\theta} \ll Lc_d/c^{1/2}c_{d-1}\,,
\end{eqnarray}
which gives, in $d=3$
\begin{eqnarray}
v\gg \frac{TK_0^2}{2\pi Lc} \,.
\end{eqnarray}
In the limit $L\to\infty$, we see that this last condition is satisfied nearly everywhere in the interval $v\in[0,1]$.
For films of finite thickness, one can easily show that equations (\ref{approx-sigma})-(\ref{solution-sigma}) are still valid
provided that the thickness $L$ satisfies
\begin{eqnarray}
L \gg \frac{TK_0^2}{2\pi c} \,.
\end{eqnarray}
a condition which is satisfied by most thin HTSC films even for temperatures close to the superconducting critical temperature $T_c$.

The above solution (\ref{solution-sigma}) for $[\sigma](v)$ is a priori valid up to a critical value $v_c$, above 
which $[\sigma](v)$ is just a constant, $[\sigma](v)=(v_c/v_0)^{2/\theta}$.
With the knowledge of the analytic form of the function $[\sigma](v)$, we now are in a position to 
find the diagonal part $\tilde{G}({\bf q},q_n)$ such that 
\begin{eqnarray}
\tilde{C}({\bf x},z) =  \frac{2T}{L}\sum_{n=-\infty}^\infty\int\frac{d^{d-1}\bf q}{(2\pi)^{d-1}}
\big(1-\cos({\bf q}\cdot{\bf x}+q_nz)\big)\,\tilde{G}({\bf q},q_n) \,,
\end{eqnarray}
and which is given by:\cite{Mezard,Giamarchi1}
\begin{eqnarray}
\tilde{G}({\bf q},q_n) = \frac{1}{c(q^2+q_n^2)}\Big( 1 + 
\int_0^1\frac{dv}{v^2}\frac{[\sigma](v)}{c(q^2+q_n^2) + [\sigma](v)}\Big)\,.
\end{eqnarray}
Specializing to the three-dimensional case $(d=3)$, if we now
use the result (\ref{solution-sigma}) with $\theta=d-2=1$ and change the variable of integration to 
$x=v^{2/\theta}=v^2$, we obtain 
\begin{eqnarray}
\int_0^{v_c} \!\!\frac{dv}{v^2}\frac{[\sigma](v)}{c(q^2+q_n^2) + [\sigma](v)}\! = 
\!\frac{1}{2}\int_0^{v_c^2}\!\!\frac{dx}{\sqrt{x}\big(x + cv_0^2(q^2+q_n^2)\big)} \,.
\end{eqnarray}
Now, using the result\cite{Prudnikov}
$$\int\frac{dx}{\sqrt{x}(x+a)} = \frac{2}{\sqrt{a}}\,\,\mbox{Arctan}(\sqrt{x/a})\,,$$
we finally obtain
\begin{eqnarray}
\int_0^{v_c} \frac{dv}{v^2}\;\frac{[\sigma](v)}{c(q^2+q_n^2) + [\sigma](v)}  = 
\frac{1}{\sqrt{cv_0^2}}\,\frac{1}{\sqrt{q^2+q_n^2}}\,
\mbox{Arctan}\Big(v_c/\sqrt{cv_0^2(q^2+q_n^2)}\,\Big).
\end{eqnarray}
On the other hand, using the result $[\sigma](v)=(v_c/v_0)^2$ for $v_c<v<1$ it is not difficult to see that
\begin{eqnarray}
\int_{v_c}^1\frac{dv}{v^2}\frac{[\sigma](v)}{c(q^2+q_n^2)+[\sigma](v)} = 
\frac{\big(\frac{1}{v_c}-1\big)(v_c/v_0)^2}{c(q^2+q_n^2) +(v_c/v_0)^2}.
\end{eqnarray}
Using the fact that $\sqrt{cv_0^2}=TK_0^2/4\pi c$, and approximating 
$$\mbox{Arctan}\left(\frac{4{\pi}c v_c }{TK_0^2\sqrt{q^2+q_n^2} }\right)\simeq \pi/2\,,$$ 
(as we are mainly interested in the $({\bf q},\,q_n)\to 0$ behavior of the correlation function), 
we find, in the long-wavelength limit
\begin{eqnarray}
\tilde G({\bf q},q_n) \approx \Big(\frac{2\pi^2}{TK_0^2}\Big)\,\frac{1}{|q^2+q_n^2|^{3/2}} \,,
\label{result-G(q)}
\end{eqnarray}
which is nothing but equation (3.18) of reference\cite{Giamarchi1} in $d=3$. 
The purpose of the calculation above, equations (\ref{exp-sigma}) to (\ref{result-G(q)}),
was to show that nothing actually changes in Giamarchi and Le Doussal's derivation  
when we use the decomposition of the $q_z$ modes into center of mass and internal modes as
we have done throughout this paper, and that the $1/q^d$ behavior of $\tilde G({\bf q},q_n)$ in 
$d$ dimension
previously found by these authors is still valid here.
From expression (\ref{result-G(q)}), we can easily find
the expression of the correlation function $\tilde{C}({\bf x},z)$.
We have:
\begin{eqnarray}
\tilde{C}({\bf x},z) = \tilde{C}_0({\bf x}) + \tilde{C}_1({\bf x},z)\,,
\end{eqnarray}
where 
\begin{eqnarray}
\tilde{C}_0({\bf x}) = \frac{2T}{L}\int\frac{d^{d-1}\bf q}{(2\pi)^{d-1}}
\big(1-\cos({\bf q}\cdot{\bf x}+q_nz)\big)\tilde{G}({\bf q},0) \,,
\end{eqnarray}
is restricted to the CM mode, while
\begin{eqnarray}
\tilde{C}_1({\bf x},z) & = & \frac{2T}{L}\sum_{n\neq 0}\int\frac{d^{d-1}\bf q}{(2\pi)^{d-1}}
\big(1-\cos({\bf q}\cdot{\bf x}+q_nz)\big)\,\tilde{G}({\bf q},q_n) \,,
\nonumber\\
& \simeq & \frac{2T}{L}\int\frac{d^d\bf q}{(2\pi)^d}
\big(1-\cos({\bf q}\cdot{\bf x}+q_zz)\big) \tilde{G}({\bf q},q_n)\,
\end{eqnarray}
corresponds to the internal modes of the flux lines and
is the correlation function calculated by Giamarchi and Le Doussal.\cite{Giamarchi1}
We have, for the CM correlation function $\tilde{C}_0({\bf x})$~:
\begin{eqnarray}
\tilde{C}_0({\bf x}) & = & \frac{4\pi^2T}{LTK_0^2}
\,\int\frac{d^2\bf q}{(2\pi)^2}\,\,\frac{1-\cos({\bf q}\cdot{\bf x})}{q^3} \,,
\nonumber\\
& = & \frac{2\pi }{LK_0^2}\int_0^\Lambda dq\,\,\frac{1-J_0(qx)}{q^2}\,.
\end{eqnarray}
Changing variables from $|{\bf x}|$ to $u=\Lambda|{\bf x}|$, we obtain
\begin{eqnarray}
\tilde{C}_0({\bf x}) = \Big(\frac{2\pi}{LK_0^2}\Big)\,|{\bf x}|
\,\times\,\int_0^{\Lambda |{\bf x}|} du \,\,\frac{1-J_0(u)}{u^2}
\nonumber
\end{eqnarray}
For large values of $|{\bf x}|$, we can to a very good approximation extend the range of integration to infinity. 
Then, using the fact that\cite{Abramowitz}
$\int_0^\infty du \;\big(1-J_0(u)\big)/u^2 = 1$, we finally obtain 
\begin{eqnarray}
\tilde{C}_0({\bf x}) \simeq \Big(\frac{2\pi}{LK_0^2}\Big)\,\,|{\bf x}| \,.
\label{B0-CM}
\end{eqnarray}
On the other hand, if we use the result\cite{Giamarchi1} for $C_1({\bf x})$ (in $d=3$ dimensions)
\begin{eqnarray}
\tilde{C}_1({\bf x}) = \frac{2}{K_0^2}\ln|{\bf x}| \,,
\end{eqnarray}
we finally obtain
\begin{eqnarray}
\tilde C({\bf x},z) = \Big(\frac{2\pi}{LK_0^2}\Big)\,\,|{\bf x}| + \frac{2}{K_0^2}\ln|{\bf x}| \,.
\label{tildeB}
\end{eqnarray}
Now, the translational order correlation function at reciprocal wavevector ${\bf K}$
\begin{eqnarray}
\Psi_{K}({\bf x}) = \langle\rho^*_{\bf K}({\bf x})\rho_{\bf K}({\bf {\bf x}})\rangle
\end{eqnarray} 
is given by\cite{Giamarchi1}
\begin{eqnarray}
\Psi_{K}({\bf x}) = \mbox{e}^{-\frac{1}{2}K^2{\tilde C}({\bf x})} \,,
\end{eqnarray}
which, given the result (\ref{tildeB}), implies that $C_{K}({\bf x})$ behaves for large $|{\bf x}|$ as
\begin{eqnarray}
\Psi_{K}({\bf x}) \sim  \exp\Big(-\frac{\pi d_{\perp} K^2}{K_0^2}\;\frac{|{\bf x}|}{L}\Big)
\;\big|{\bf x}\big|^{-\frac{d_\perp \,K^2}{2K_0^2}} \,,
\label{CK-CM}
\end{eqnarray}
and hence we see that point disorder does destroy long range order of the FLL, although 
it does so only on asymptotic length scales $|{\bf x}|\sim L$.
The same conclusion would follow from more sophisticated functional renormalization-group 
arguments\cite{DSFisher,Balents-Fisher,Emig}.

Equations (\ref{B0-CM}) and (\ref{CK-CM}) are the most important results of this paper. 
They show clearly that the QLRO of the flux line lattice 
predicted by considering the full correlation function $C({\bf x},z)$ and integrating over all internal modes, is 
actually lost when one carefully separates out the fluctuations of the CM positions of the flux lines. 
As such, this result confirms our claim that the CM mode plays an important role and is in fact the most relevant 
one to look at when  considering positional order of flux line lattices in samples of highly anisotropic shapes $L\ll L_\perp$.
For such samples, we expect the destruction of long range range order to be 
observable in experiments.

\section{Positional order in the moving flux line lattice in the presence of disorder}
\label{lro-moving}

The considerations of the past section can be generalized to the case of a moving FLL in a disordered potential. Here,
we shall restrict ourselves to the determination of the dynamic Larkin lengths\cite{Giamarchi3} for the center of mass
positions. Following standard arguments
\cite{Sneddon-Cross-Fisher,BalentsMPAFisher,Giamarchi3,BMR,Scheidl-Vinokur}, 
one can show that the usual perturbative
expression for the mean-square displacement of a FLL drifting with mean velocity $v$ along the ${\bf x}$ direction 
and subject to a random force ${\bf F}({\bf r})$ with the correlations (\ref{randomforce});
\begin{eqnarray}
C({\bf r}) = 2W\int\frac{d^d{\bf q}}{(2\pi)^d}\,\,
\frac{1-\cos({\bf q}\cdot{\bf r})}{(\eta vq_x)^2+c^2q^4} \,,
\end{eqnarray}
for a sample of finite thickness $L$ should be replaced by
\begin{eqnarray}
C({\bf r}) = \frac{2W}{L}\sum_n\int\frac{d^{d-1}{\bf q}}{(2\pi)^{d-1}}\,\,
\frac{1-\cos({\bf q}\cdot{\bf x}+q_nz)}{(\eta vq_x)^2+c^2(q^2+q_n^2)^2} \,.
\end{eqnarray}
In the above expressions, $\eta$ is the microscopic friction coefficient and isotropic elasticity with 
an elastic constant
$c$ is assumed. The mean square displacement of the center of mass mode is given by the $n=0$ term in the above sum,
namely
\begin{eqnarray}
C_0({\bf x}) = \frac{2W}{L}\int_{q<\Lambda}\frac{d^{d-1}\bf q}{(2\pi)^{d-1}}\,\,
\frac{1-\cos({\bf q}\cdot{\bf x})}{(\eta vq_x)^2 + c^2q^4} \,.
\label{ms-cm}
\end{eqnarray}
The integral in (\ref{ms-cm}) has been caculated by several authors\cite{RTcdw,Giamarchi3,BMR}. 
We obtain:
\begin{eqnarray}
C_0(x,{\bf y}) = \frac{2W |y|^{4-d}}{\eta L cv}\,F(cx/\eta vy^2) \,,
\end{eqnarray} 
where ${\bf y}$ represents all the space variables other than $x$ (the direction of the drive) and $z$ (the direction
of the magnetic field ${\bf B}$), and where
the scaling function $F$ is such that $F(0)=\mbox{const}$ and $F(x)\sim x^{(3-(d-1))/2}\sim x^{(4-d)/2}$ when
$|x|\gg 1$. Defining the dynamical Larkin lengths for the center of mass displacements $R_0^x$ and $R_0^y$ by
the equation $C(R_0^x,R_0^y)\simeq\xi^2$, we obtain:
\begin{eqnarray}
R_0^y & \simeq & \Big(\frac{\eta c v L\xi^2}{2W}\Big)^{1/(4-d)} \,,
\\
R_0^x & \simeq &  \eta v (R_0^y)^2/c \,.
\end{eqnarray}
Specializing to the case $d=3$, we see that $R_0^y$ grows only linearly with $v$, in contrast to the transverse Larkin
length of the total fluctuations\cite{RTcdw,Giamarchi3,BMR,Scheidl-Vinokur}
which exhibits exponential growth as a function of $v$, $R_c^y\sim \exp(\eta vc\xi^2)/\Lambda$.
The critical lengths $(R_0^x,R_0^y)$ obtained here can however be quite large for macroscopic samples with a large
thickness $L$.

\section{Conclusion}
\label{lro-conc}

In this article, we have reexamined the problem of the positional order of the Abrikosov flux 
line lattice in type II superconducting samples of finite thickness $L$, carefully separating 
the center of mass mode of the flux lines from their internal modes. 
While this separation turns out to be unimportant for clean systems 
with thermal fluctuations, in the presence of a random pinning potential we find that the Larkin 
length governing the growth of fluctuations of the center of mass positions is very large for 
macroscopic samples and grows like $\sqrt{L}$ in $2+1$ dimensions, which suggests that flux line 
lattices on average retain translational order on length scales much larger than the lengths 
predicted so far (the latter corresponding to the growth of the internal modes fluctuations of 
the lines). Going beyond the simple Larkin analysis, within a Gaussian variational approximation 
with broken replica symmetry we find that translational order of the three dimensional flux line 
lattice is destroyed, and that the logarithmic roughness predicted in previous work crosses over 
to power-law growth of the vortex displacements. Although this destruction of positional 
order only takes place on lateral length scales of order the sample thickness $L$, it can lead to 
experimentally measurable effects in superconducting thin films.

\appendix

\section{
Eigenvalue of the replicon mode and stability of the replica symmetric solution}
\label{lro-appendix}

In this appendix we consider the eigenvalue $\lambda$ of the so-called ``replicon'' mode, which for our 
problem is given by:\cite{Mezard,Giamarchi1}
\begin{eqnarray}
\lambda  =  1 - \frac{1}{d_\perp}\sum_{\bf K} K^4\Delta_K\mbox{e}^{-\frac{TK^2}{L}
\sum_n\int_{\bf p}G_c({\bf p},p_n)} 
\times\frac{1}{L}\sum_n\int_{\bf q} G_c^2({\bf q},q_n)
,\label{replicon}
\end{eqnarray}
where the connected propagator $G_c({\bf q},q_n)$ is defined by $G_c({\bf q},q_n)=\sum_bG_{ab}({\bf q},q_n)$. 
Using the saddle 
point equations (\ref{var-1})-(\ref{var-2}), 
it is not difficult to see that $G_c$ is given by $G_c({\bf q},q_n)=1/(cq^2+cq_n^2)$.
Here, we shall take $G_c({\bf q},q_n)=1/(cq^2+cq_n^2+\mu^2)$, with a small regularizing mass $\mu^2$,
the limit $\mu\to 0$ being taken at the end of the calculation.
Transforming the sum over modes $(1/L)\sum_n$ into $q_z$ integrals in equation (\ref{replicon}) right away leads 
to the conclusion that in $d=2+1$ dimensions the eigenvalue of the replicon mode is always negative, and hence 
that three dimensional flux line lattices are always unstable to point disorder. Here, we shall instead be 
careful to separate the CM from the internal modes, upon which we obtain (here we only consider the case $d=2+1$ 
dimensions):
\begin{eqnarray}
\sum_m\int_{\bf p} G_c({\bf p},p_m)  =   \frac{1}{4\pi c}
\ln\Big(1+\frac{\Lambda^2c}{\mu^2}\Big) +
\frac{1}{2\pi c}\sum_{m=1}^\infty\ln\Big(1+\frac{\Lambda^2}{p_m^2+\mu^2/c}\Big) \,,
\end{eqnarray}
and so, in the limit $\mu\to 0$ we can write:
\begin{eqnarray}
\frac{1}{L}\sum_m\int_{\bf p} G_c({\bf p},p_m)\big|_{\mu\to 0} =
\frac{1}{4\pi Lc}\ln\big(\frac{\Lambda^2c}{\mu^2}\big) + C_1 \,,
\end{eqnarray}
where $C_1$ is the constant given by 
\begin{equation}
C_1\simeq \frac{1}{2\pi c}\int_0^\infty\frac{dq_z}{2\pi}\ln(1+\Lambda^2/q_z^2)\,.
\end{equation}
On the other hand, one can also easily show that
\begin{eqnarray}
\frac{1}{L}\sum_m\int_{\bf p} G_c^2({\bf p},p_m)\big|_{\mu\to 0} =  \frac{1}{4\pi Lc^2\mu^2} 
+C_2\,,
\end{eqnarray}
where
\begin{equation}
C_2 = -\frac{1}{4\pi Lc^2\Lambda^2} + 
\frac{1}{2\pi Lc^2}\sum_{n\ge 1}\Big(\frac{1}{q_n^2}-\frac{1}{q_n^2+\Lambda^2}\Big)\,.
\end{equation}
Collecting all terms, we see that $\lambda$ can be written as
\begin{eqnarray}
\lambda = 1 - \frac{1}{d_\perp}\sum_{\bf K} K^4\Delta_K\mbox{e}^{-TK^2C_1}
\Big(\frac{\Lambda^2c}{\mu^2}\Big)^{-TK^2/4\pi Lc} \times
\big[\frac{1}{4\pi Lc^2\mu^2} + C_2\big] \,.
\end{eqnarray}
In the limit $\mu\to 0$, the dominant term in the above sum is the term with $K=K_0$, which 
gives:
\begin{eqnarray}
\lambda \approx 1 - \frac{K_0^4\Delta_{K_0}}{4{\pi}Ld_{\perp}c^2}\mbox{e}^{-TK_0^2C_1}
(\Lambda^2c)^{-\frac{TK_0^2}{4\pi Lc}}
\mu^{-2\big(1- \frac{TK_0^2}{4\pi Lc}\big)}\,.
\end{eqnarray}
From this last expression, we see that the sign of $\lambda$ when $\mu\to 0$ depends on the value of 
$TK_0^2/4\pi Lc$~:

$\bullet$ For $TK_0^2/4\pi Lc>1$, i.e. $T>T_c=4\pi Lc/K_0^2$, $\lambda(\mu\to 0) = 1 > 0$, the replica-symmetric
solution is stable; 

$\bullet$ For $TK_0^2/4\pi Lc<1$, or $T<T_c=4\pi Lc/K_0^2$, $\lambda(\mu\to 0) \to -\infty$ and the replica
symmetric solution is unstable.

An identical result has been obtained by Giamarchi and Le Doussal in ref.\cite{Giamarchi2} (where the authors 
adopted a separation of modes similar to ours) in the context of correlated 
disorder. We thus see that the conclusions of this last reference regarding the stability of the replica 
symmetric solution result solely from the separation of modes into CM 
and internal modes, and has nothing to do with the nature of the (point-like or correlated) disorder considered.


\begin{thebibliography}{99}
\bibitem{footnote1}
This constant actually depends on the cut-off procedure. For a square lattice, it is
given by $\gamma+\frac{1}{2}\ln 8$, where $\gamma\simeq 0.5772$ the Euler-Mascheroni constant
(see ref.\cite{Chaikin-Lubensky}).
\bibitem{footnote2}
Note that the Hamiltonian (\ref{Htotal}) can be diagonalized exactly without minimization
(using replicas to average over the random field),
in which case one obtains an additional, subdominant contribution to the correlation functions
in equations (\ref{uu-dis-larkin}) and (\ref{newcorr-larkin}). Here we neglect this thermal part
and focus on the dominant random force contributions.
\bibitem{footnote3}
For further details on how this parametrization emerges in the replica-broken phase in
the limit $n\to 0$, we refer the reader to\cite{Dotsenko} and references therein.
\end{thebibliography}
\end{document}